\documentclass[letterpaper]{article}

\usepackage{amsmath}
\usepackage{amsfonts}
\usepackage{amsthm}
\usepackage[letterpaper]{geometry}
\usepackage{hyperref}

\DeclareMathOperator{\ex}{ex}

\title{On Unit Distances in a Convex Polygon}
\author{Amol Aggarwal}

\begin{document}

\maketitle

\begin{abstract}
In 1959, Erd\H{o}s and Moser asked for the maximum number of unit distances that may occur among the vertices of a convex $n$-gon. Until now, the best known upper bound has been $2\pi n \log_2 n + O(n)$, achieved by F\"{u}redi in 1990. In this paper we examine two properties that any convex polygon must satisfy and use them to prove several facts related to the above question. In particular, we improve upon F\"{u}redi's result, obtaining a bound of $n \log_2 n + O(n)$; we exhibit a new class of ``cycles" formed by unit distances that are forbidden in convex polygons; and we provide a lower bound that shows the limitations of our methods. The second result answers a question of Fishburn and Reeds in the negative.
\end{abstract}

\newtheorem{thm}{Theorem}

\section{Introduction}
\subsection{Background}
Let $d(v, u)$ denote the Euclidean distance between any two points $v$ and $u$ in the plane. If $d(v, u) = 1$, we say that $v$ and $u$ form a {\itshape unit distance} and we call the edge $vu$ a unit distance. For any set of points $S$ in the plane, let $U(S)$ denote the number of unit distances formed among the elements of $S$. In 1946, Erd\H{o}s asked for the value of $U(n) = \max U(S)$, where $S$ ranges over all sets of $n$ points in the plane [4]. He conjectured that $U(n) = o(n^{1 + \varepsilon})$ for any $\varepsilon > 0$ and proved that $U(n) \le n^{3/2}$ using the fact that any two unit circles may have at most two points in common. In [11], Spencer, Szemer\'{e}di and Trotter improved this bound to $U(n) = O(n^{4/3})$. 

In 1959, Erd\H{o}s and Moser posed a variant of the original question. Identifying convex polygons with their vertex sets, they asked for the value of $U_c (n) = \max U(\mathcal{P})$, where $\mathcal{P}$ ranges over all convex $n$-gons in the plane [5]. They conjectured that $U_c (n) = \Theta (n)$ and showed that $U_c (n) \ge \lfloor 5(n-1)/3 \rfloor$. In [3], Edelsbrunner and Hajnal improved the lower bound to $U_c (n) \ge 2n - 7$. The upper bound on $U_c (n)$ was improved from $O(n^{4 / 3})$ to $2 \pi n \log_2 n + O(n)$ by F\"{u}redi, who used $0-1$ matrices to represent convex polygons [7]. 

Let us describe the relationship between $0-1$ matrices and unit distances. Different authors have done this in different ways; in what follows, we outline a variant of the method given by Fishburn and Reeds in [6]. A real matrix is a matrix with real entries and a $0-1$ matrix is a matrix whose entries are either $0$ or $1$. Observe that a $0-1$ matrix may be recovered from a real matrix by replacing all entries not equal to $1$ with $0$; the resulting $0-1$ matrix is called the {\itshape skeleton} of the original matrix. 

For any convex $n$-gon $\mathcal{P} = v_1 v_2 \ldots v_n$, vertices listed in clockwise order, we say that two vertices $v_i$ and $v_j$ (with $1\le i < j \le n$) are {\itshape antipodal} with respect to $\mathcal{P}$ if there exist parallel lines $l_1$ through $v_i$ and $l_2$ through $v_j$ such that $\mathcal{P}$ is contained in the strip of the plane bounded by $l_1$ and $l_2$. Taking indices modulo $n$, consider the convex polygons $\mathcal{P}_1 = v_i v_{i + 1} \ldots v_{j - 1}$ and $\mathcal{P}_2 = v_j v_{j + 1} \ldots v_{n + i - 1}$. The partition $\mathcal{P} = \mathcal{P}_1 \cup \mathcal{P}_2$ is called an {\itshape antipodal cut} of $\mathcal{P}$. We remark that Pach and Brass used antipodal cuts in [2] to inductively prove that $U_c (n) \le 9.65 n \log_2 n$. 

Let $a = j - i$ and $b = n - a$; relabel the vertices of $\mathcal{P}_1$ and $\mathcal{P}_2$ by setting $u_k = v_{i + k - 1}$ for $1\le k \le a$ and $w_k = v_{i - k}$ for $1\le k \le b$. Consider the $a \times b$ {\itshape distance matrix} $\textbf{D}_{\mathcal{P}} = \textbf{D}_{\mathcal{P}, i, j}$ whose $(r, c)$ entry is $d(u_r, w_c)$ for each $(r, c) \in [1, a] \times [1, b]$. Let $\textbf{M}_{\mathcal{P}}$ be the skeleton of $\textbf{D}_{\mathcal{P}}$; we call $\textbf{M}_{\mathcal{P}}$ a {\itshape $0-1$ cut matrix} associated with $\mathcal{P}$. Let $U(\textbf{M}_{\mathcal{P}})$ denote the number of entries in $\textbf{M}_{\mathcal{P}}$ equal to $1$; then $U(\textbf{M}_{\mathcal{P}})$ is equal to the number of unit distances $uw$ with $u \in \mathcal{P}_1$ and $w \in \mathcal{P}_2$. This implies that $U(\mathcal{P}) = U(\mathcal{P}_1) + U(\mathcal{P}_2) + U(\textbf{M}_{\mathcal{P}})$. It may be shown (see \hyperref[antipodalcut]{Proposition \ref*{antipodalcut}} of Section 2) that $U(\mathcal{P}_1) + U(\mathcal{P}_2) \le 2n$; thus, showing $U_c (n) = \Theta (n)$ amounts to proving $U(\textbf{M}_{\mathcal{P}}) = O(n)$.

Several authors have attempted to prove this upper bound through the use of forbidden matrices. Let $\textbf{A} = \{ a_{i, j} \}$ be an $n_1 \times n_2$ real matrix; we call a $k_1 \times k_2$ real matrix $\textbf{B} = \{ b_{i, j} \}$ a {\itshape submatrix} of $\textbf{A}$ if there are integers $1 \le i_1 < i_2 < \cdots < i_{k_1} \le n_1$ and $1\le j_1 < j_2 < \cdots < j_{k_2} \le n_2$ such that $b_{r, c} = a_{i_r, j_c}$ for each $(r, c) \in [1, k_1] \times [1, k_2]$. Following Tardos in [12], we say that $\textbf{A}$ {\itshape contains} $\textbf{B}$ if there exist integers $i_1, i_2, \ldots , i_{k_1}$ and $j_1, j_2, \ldots , j_{k_2}$ as above such that $b_{r, c} = 1$ implies that $a_{i_r, j_c} = 1$ for each $(r, c)$; otherwise $\textbf{A}$ {\itshape avoids} $\textbf{B}$. We call a real matrix {\itshape forbidden} if any $0-1$ cut matrix avoids it. Examples of known (see [6]) forbidden matrices are shown below 

\begin{center}
$\textbf{S}_2 = \textbf{T}_2 = \left[ \begin{array}{cc} 1 & 1 \\ 1 & 1 \end{array} \right]; \quad \textbf{G} = \left[ \begin{array}{ccc} 1 & 1 \\ & & 1 \\ 1 & & 1\end{array} \right]; \quad \textbf{H} = \left[ \begin{array}{ccc} 1 & & 1 \\ 1 \\ & 1 & 1 \end{array} \right]$ \\

$\textbf{S}_3 = \left[ \begin{array}{ccc}  & 1 & 1 \\ 1 & 1 \\ 1 & & 1 \end{array} \right]; \quad \textbf{T}_3 = \left[ \begin{array}{ccc} 1 & & 1 \\ & 1 & 1 \\ 1 & 1\end{array} \right]; \quad \textbf{C} = \left[ \begin{array}{cccc} 1 & 1 \\ 1 \\ & & & 1 \\ & & 1 & 1 \end{array} \right]$ \\

$\textbf{C}_1 = \left[ \begin{array}{cccc} 1 & & 1\\ 1 \\ & & & 1 \\ & 1 & & 1\end{array} \right]; \quad \textbf{C}_2 = \left[ \begin{array}{cccc} 1 & 1 \\ & & & 1 \\ 1 \\ & & 1 & 1\end{array} \right]; \quad \textbf{C}_3 = \left[ \begin{array}{cccc} 1 & & 1 \\ & & & 1 \\ 1 \\ & 1 & & 1\end{array} \right]$, 
\end{center}

\noindent where the blank entries denote zeroes. We remark that Brass, K\'{a}rolyi, and Valtr used the fact that $\textbf{C}$ is forbidden to show that $U_c (n) \le 7n \log_2 n$ in [1]. 

Observe that $\textbf{S}_2$, $\textbf{S}_3$, and $\textbf{T}_3$ are elements of a larger class of matrices known as staircase matrices. For any integer $n$, let $\textbf{S}_n = \{ s_{i, j} \}$ denote the $n \times n$ $0-1$ matrix satisfying $s_{n, n} = s_{1, n} = s_{i, n - i} = s_{i + 1, n - i} = 1$ for all integers $1\le i\le n - 1$ and $s_{i, j} = 0$ for all other $i$ and $j$. For any integer $n$, let $\textbf{T}_n = \{ t_{i, j} \}$ denote the $n \times n$ $0-1$ matrix satisfying $t_{1, 1} = t_{n, 1} = t_{i, n - i + 1} = t_{i + 1, n - i + 1} = 1$ for all integers $1\le i\le n - 1$ and $t_{i, j} = 0$ for all other $i$ and $j$. The set of {\itshape staircase matrices} is the union $\bigcup_{i = 2}^{\infty} \{ \textbf{S}_i, \textbf{T}_i \}$. Following Fishburn and Reeds, we call a $0-1$ matrix {\itshape pattern feasible} if it avoids each of the nine matrices above as well as each staircase matrix; it may be shown that if a $0-1$ matrix is not pattern feasible, then it is forbidden [6]. In the same paper Fishburn and Reeds asked whether any pattern feasible matrix is in fact a $0-1$ cut matrix. 

\subsection{Results}

Our first result is an improvement upon F\"{u}redi's bound on $U_c (n)$ by a multiplicative factor of $2 \pi$; we achieve this by combining various results about $0-1$ matrices (see Sections 2.1 and 2.2).

\begin{thm}
\label{upperbound}
For each positive integer $n$, $U_c (n) \le n\log_2 n + 4n$. 
\end{thm}

Our next two results answer and generalize the question asked by Fishburn and Reeds. The novel aspect of this paper that allows us to accomplish this is our analysis of the distance matrix, which contains more refined information than does the $0-1$ cut matrix. We will use two properties, which we call the diagonal property and the obtuse angle property, in order to perform this analysis. Before stating our remaining results, let us introduce these two properties and describe their relationship with pattern feasible matrices. 

A real matrix has the {\itshape diagonal property} if it has positive entries and has no $2 \times 2$ submatrix $\textbf{M} = \{ m_{i, j} \}$ satisfying $m_{1, 1} + m_{2, 2} \ge m_{1, 2} + m_{2, 1}$. \hyperref[diagonalproperty]{Proposition \ref*{diagonalproperty}} of the next section states that any distance matrix has the diagonal property. 

For any $2 \le d, e \le 4$, a $d \times e$ real matrix $\textbf{M} = \{ m_{i, j} \}$ is called an {\itshape acute angle matrix} if there are integers $ r_1, \in [2, d]$, $c_1 \in [2, e]$, $r_2 \in [1, d - 1]$, and $c_2 \in [1, e - 1]$ such that $m_{1, 1} \ge m_{1, c_1}, m_{r_1, 1}$ and $m_{d, e} \ge m_{r_2, e}, m_{d, c_2}$. For instance, any real matrix whose skeleton is one of the matrices $\{ \textbf{S}_2, \textbf{G}, \textbf{H}, \textbf{C}, \textbf{C}_1, \textbf{C}_2, \textbf{C}_3 \}$ from Section 1.1 is an acute angle matrix; however, neither $\textbf{S}_3$ nor $\textbf{T}_3$ is an acute angle matrix. A matrix with positive entries that has no acute angle submatrix is said to have the {\itshape obtuse angle property}; \hyperref[obtuseangleproperty]{Proposition \ref*{obtuseangleproperty}} of the next section states that any distance matrix has the obtuse angle property. 

We remark that the diagonal property has been used by Pach and Tardos in [10] to obtain another proof of the bound $U(n) = O(n^{4 / 3})$. Some specific cases of the obtuse angle property have also been discussed in previous works such as [1] and [6]. However, we have not seen it used in generality until now. 

A real matrix that has both the diagonal property and the obtuse angle property is called a {\itshape distance-like matrix}; any distance matrix is distance-like. One may verify that any distance-like matrix has a pattern feasible skeleton. However, there exists a pattern feasible matrix that is not the skeleton of any distance-like matrix. Let us describe a class of real matrices that contains such an element. 

Suppose that $k_1$ and $k_2$ are integers greater than $1$. A $k_1 \times k_2$ real matrix $\textbf{M} = \{ m_{i, j} \}$ is a {\itshape cycle with an intersection-free edge} if there exist positive integers $r_1 = 1; r_2, \ldots , r_l \ne 1$ less than or equal to $k_1$ and $c_1 = 1; c_2, \ldots , c_l \ne 1$ less than or equal to $k_2$ such that $r_i \ne r_{i + 1}$ and $c_i \ne c_{i + 1}$ for each $1\le i\le l - 1$ and such that $m_{r_i, c_i} = 1 = m_{r_i, c_{i + 1}}$ for each $1\le i\le l$, where indices are taken modulo $l$. For instance, the staircase matrix $\textbf{T}_k$ is a cycle with an intersection-free edge for any integer $k \ge 2$. Moreover, any real matrix whose skeleton is the pattern feasible matrix $\textbf{E}$ below is a cycle with an intersection-free edge. 

\begin{center}
$\textbf{E} = \left[ \begin{array} {cccc} 1 & & &1 \\ &1 &1  \\ &1 & &1\\ 1 & &1  \end{array} \right]$
\end{center}

\noindent We have the following result on cycles with an intersection-free edge. 

\begin{thm}
\label{cycle}
No cycle with an intersection-free edge is a distance-like matrix. 
\end{thm}

In particular, Theorem 2 implies that there is no distance-like matrix whose skeleton is $\textbf{E}$; thus the pattern feasible matrix $\textbf{E}$ is not a $0-1$ cut matrix. This yields a negative answer to the question posed by Fishburn and Reeds. 

Our final result shows the limitations of the diagonal and obtuse angle properties; they alone will not suffice to obtain $U_c (n) = \Theta (n)$. 

\begin{thm}
\label{distancelike}
For any positive integer $m$, there exists a $2^m\times 2^m$ distance-like matrix with $2^{m-1}(m + 1)$ entries equal to $1$.
\end{thm}

\section{Proofs of Theorems 1, 2, and 3}

\subsection{Preliminary Facts}

In this subsection we collect several facts that will be used later in the article. The first fact is about antipodal cuts and is due to Brass and Pach [2]. 

\newtheorem{prop}{Proposition}
\newtheorem{lem}{Lemma}

\begin{prop}
\label{antipodalcut}
Let $\mathcal{P}$ be a convex $n$-gon and let the partition $\mathcal{P} = \mathcal{P}_1 \cup \mathcal{P}_2$ be an antipodal cut. Then $U(\mathcal{P}_1) + U(\mathcal{P}_2) \le 2n$. 
\end{prop}

The next two facts together show that any distance matrix is distance-like. A proof of \hyperref[diagonalproperty]{Proposition \ref*{diagonalproperty}} is given in [10], and a special case of \hyperref[obtuseangleproperty]{Proposition \ref*{obtuseangleproperty}} is used in [1]. 

\begin{prop}
\label{diagonalproperty}
Any distance matrix satisfies the diagonal property. 
\end{prop}

\begin{prop}
\label{obtuseangleproperty}
Any distance matrix satisfies the obtuse angle property. 
\end{prop}

\begin{proof} 
Suppose to the contrary that there exist convex polygons $\mathcal{P}$, $\mathcal{P}_1 = u_1 u_2 \ldots u_a$, and $\mathcal{P}_2 = w_1 w_2 \ldots w_b$ such that the partition $\mathcal{P} = \mathcal{P}_1 \cup \mathcal{P}_2$ is an antipodal cut and such that the distance matrix $\textbf{D}_{\mathcal{P}}$ associated with this cut does not have the obtuse angle property. By replacing it with one of its acute angle submatrices if necessary, we will assume that $\textbf{D}_{\mathcal{P}}$ is an acute angle matrix. Then there exist integers $r_1 \in [2, a]$, $c_1 \in [2, b]$, $r_2 \in [1, a - 1]$, and $c_2 \in [1, b - 1]$ such that $d(u_1, w_1) \ge d(u_{r_1}, w_1), d(u_1, w_{c_1})$ and $d(u_a, w_b) \ge d(u_{r_2}, w_b), d(u_a, w_{c_2})$. The first inequality implies that $\angle w_1 u_1 u_{r_1} \le \angle w_1 u_{r_1} u_1$, so $\angle w_1 u_1 u_a \le \angle w_1 u_1 u_{r_1} < \pi / 2$. All other angles of the quadrilateral $w_1 u_1 u_a w_b$ are acute by similar reasoning; this is a contradiction. 
\end{proof}

The next two facts are about $0-1$ matrices. Before stating these facts, let us define some relevant terminology. For positive integers $a$ and $b$ and a $0-1$ matrix $\textbf{M}$, let $\ex(a, b, \textbf{M})$ denote the maximum number of entries equal to $1$ in an $a \times b$ $0-1$ matrix avoiding $\textbf{M}$. 

For an $r_1 \times c_1$ $0-1$ matrix $\textbf{M} = \{ m_{i, j} \}$ whose bottom-right entry equals $1$ and an $r_2 \times c_2$ $0-1$ matrix $\textbf{N} = \{ n_{i, j} \}$ whose top-left entry is equal to $1$, define the {\itshape amalgam} of $\textbf{M}$ and $\textbf{N}$ to be the $(r_1 + r_2 - 1) \times (c_1 + c_2 - 1)$ $0-1$ matrix $\textbf{L} = \{ l_{i, j} \}$ formed by attaching the bottom-right corner of $\textbf{M}$ to the top-left corner of $\textbf{N}$. Specifically, set $l_{i, j} = m_{i, j}$ for all $(i, j) \in [1, r_1] \times [1, c_1]$; $l_{i, j} = n_{i - r_1 + 1, j - c_1 + 1}$ for all $(i, j) \in [r_1, r_1 + r_2 - 1] \times [c_1, c_1 + c_2 - 1]$; and all other $l_{i, j} = 0$. The following two lemmas are due to Keszegh [8, 9] and Tardos [12], respectively.

\begin{lem}
\label{amalgam}
Let $\textbf{M}$ and $\textbf{N}$ be $0-1$ matrices whose amalgam exists and is equal to $\textbf{L}$. For all positive integers $a$ and $b$, $\ex (a, b, \textbf{L}) \le \ex (a, b, \textbf{M}) + \ex (a, b, \textbf{N})$. 
\end{lem}

\begin{lem} 
\label{exclusionnumber}
Let

\begin{center}
$\textbf{A}= \left[\begin{array} {ccc} 1  &1 \\ 1 \\ &1\end{array}\right]; \quad \textbf{B} = \left[\begin{array} {ccc} 1 & &1 \\ &1 &1\end{array}\right]$. 
\end{center}

\noindent For all positive integers $a$ and $b$, $\ex(a, b, \textbf{A}) \le (\frac{a + b}{2})\log_2 (a + b) + 2b$ and $\ex (a, b, \textbf{B}) \le (\frac{a + b}{2})\log_2 (a + b) + 2a$. 
\end{lem}

\subsection{Proof of Theorem 1}

Let $\mathcal{P}$ be a convex $n$-gon and let $\textbf{M}_{\mathcal{P}}$ be a $0-1$ cut matrix associated with $\mathcal{P}$; suppose that $\textbf{M}_{\mathcal{P}}$ has $a$ rows and $b$ columns. Define the $0-1$ matrices 

\begin{center}
$\textbf{A}= \left[\begin{array} {ccc} 1  &1 \\ 1 \\ &1\end{array}\right]; \quad \textbf{B} = \left[\begin{array} {ccc} 1 & &1 \\ &1 &1\end{array}\right]; \quad \textbf{C}' = \left[\begin{array} {cccc} 1 &1 \\ 1 \\ &1 & &1 \\  & &1 &1 \end{array} \right]$. 
\end{center}

\noindent Since no real matrix containing $\textbf{C}'$ has the obtuse angle property, \hyperref[obtuseangleproperty]{Proposition \ref*{obtuseangleproperty}} implies that $\textbf{M}_{\mathcal{P}}$ avoids $\textbf{C}'$. Hence, $U(\textbf{M}_{\mathcal{P}}) \le \ex(a, b, \textbf{C}') \le \ex(a, b, \textbf{A}) + \ex(a, b, \textbf{B})$ by \hyperref[amalgam]{Lemma \ref*{amalgam}}. By \hyperref[exclusionnumber]{Lemma \ref*{exclusionnumber}} and the fact that $a + b = n$, this quantity is at most $n \log_2 n + 2n$. It follows from \hyperref[antipodalcut]{Proposition \ref*{antipodalcut}} that $U(\mathcal{P}) \le 2n + U(\textbf{M}_{\mathcal{P}}) \le n \log_2 n + 4n$. 

\subsection{Proof of Theorem 2} 
Suppose that $\textbf{M} = \{ m_{i, j} \}$ is an $n_1 \times n_2$ cycle with an intersection-free edge. Then there exist integers $r_1 = 1; r_2, r_3, \ldots , r_l \ne 1$ and $c_1 = 1; c_2, c_3, \ldots , c_l \ne 1$ such that $r_i \ne r_{i + 1}$, $c_i \ne c_{i + 1}$, and $m_{r_i, c_i} = 1 = m_{r_i, c_{i + 1}}$ for each $1\le i\le l$, where indices are taken modulo $l$. We will deduce that $\textbf{M}$ is not distance-like by showing that $\textbf{M}$ does not have the obtuse angle property; hence it suffices to find an acute angle submatrix of $\textbf{M}$. 

Let $s$ be the minimal integer greater than $1$ such that $m_{1, s} \le 1$. The minimality of $s$ implies that $c_2 \ge s$. Let $j \in [2, l]$ be the minimal integer satisfying $c_{j + 1} < s \le c_j$; since $c_{l + 1} = c_1 = 1 < s \le c_2$, such a $j$ exists. Moreover, let $h$ be the largest integer in $[1, j - 1]$ such that $r_{h + 1} > r_h$ and let $k$ be the largest integer in $[1, j - 1]$ such that $c_{k + 1} > c_k$. The acute angle submatrix we find will depend on whether $k \le h$ or $h < k$. 

Suppose first that $k \le h$. The maximality of $h$ implies that $r_{h + 1} \ge r_j$; the maximality of $k$ and the fact that $k \le h$ implies that $c_{h + 1} > c_{h + 2}$ and that $c_{h + 1} \ge c_j \ge s$. The minimality of $j$ then yields $c_{h + 2} \ge c_{j + 1}$. Thus intersecting the first, $r_j$th, $r_h$th, and $r_{h + 1}$st rows with the $c_{j + 1}$st, $s$th, $c_{h + 2}$nd, and $c_{h + 1}$st columns yields an acute angle submatrix of $\textbf{M}$ because $m_{1, c_{j + 1}} \ge 1 \ge m_{1, s}, m_{r_j, c_{j + 1}}$ and $m_{r_{h + 1}, c_{h + 1}} = 1 = m_{r_h, c_{h + 1}}, m_{r_{h + 1}, c_{h + 2}}$. 

If $h < k$ holds instead, then intersecting the first, $r_j$th, $r_{k + 1}st$, and $r_k$th rows with the $c_{j + 1}$st, $s$th, $c_k$th, and $c_{k + 1}$st columns yields an acute angle submatrix by similar reasoning. Thus in either case, $\textbf{M}$ has an acute angle submatrix and hence cannot be distance-like. 

\subsection{Proof of Theorem 3}
We will create this distance-like matrix through a recursion. For each positive integer $m$, define the $2^m \times 2^m$ matrix $\textbf{X}_m = \{ x_{m, i, j} \}$ to satisfy $x_{m, i, j} = 0$ if $i + j = 2^m + 1$; $x_{m, i, j} = i + (2^{m + j} - i)^2 5^{-10^m}$ if $i + j > 2^m + 1$; and $x_{m, i, j} = -2^{10^m - 2i - 2j}$ otherwise. Also define the $2^m \times 2^m$ matrix $\textbf{Y}_m = \{ y_{m, i, j} \}$ to satisfy $y_{m, i, j} = -5^{10^m} ij$ for all integers $1\le i, j\le 2^m$. Recursively define the $2^m \times 2^m$ matrices $\textbf{Z}_m = \{ z_{m, i, j} \}$ through the relations

\begin{center}
$\textbf{Z}_1 = \left[ \begin{array} {cc} 1 / 4 & 0 \\ 0 & - 1 / 2 \end{array} \right]; \quad \textbf{Z}_r = 10^{-1000^r} \left[ \begin{array} {cc} \textbf{X}_{r - 1} & 10^{-1000^r} \textbf{Z}_{r - 1} \\ 10^{-1000^r} \textbf{Z}_{r - 1} & \textbf{Y}_{r - 1} \end{array} \right]$ 
\end{center}

\noindent for all integers $r \ge 2$. For each positive integer $m$, let $\textbf{D}_m$ denote the $2^m \times 2^m$ matrix formed by adding $1$ to each entry of $\textbf{Z}_m$. We claim that $\textbf{D}_m$ satisfies the conditions of Theorem 3. 

By induction on $m$, one can see that the magnitude of each entry in $\textbf{Z}_m$ is less than $1$ for each positive integer $m$. Hence each entry of $\textbf{D}_m$ is positive. Furthermore, there are $2^{r - 1}$ entries equal to $0$ in $\textbf{X}_{r - 1}$ and no such entries in $\textbf{Y}_{r - 1}$ for each integer $r \ge 2$; induction on $m$ then yields that there are $2^{m - 1} (m + 1)$ entries equal to $0$ in $\textbf{Z}_m$. Therefore, there are $2^{m - 1} (m + 1)$ entries equal to $1$ in $\textbf{D}_m$.

It remains to show that $\textbf{D}_m$ has the obtuse angle property and the diagonal property for each positive integer $m$. Let us first verify the obtuse angle property; it suffices to check that $\textbf{Z}_m$ contains no acute angle submatrix. Suppose this is false, and let $s \ge 2$ be the minimal positive integer such that $\textbf{Z}_s$ has an acute angle submatrix. Then there exist integers $i_1, i_2, i_3, i_4, j_1, j_2, j_3, j_4 \in [1, 2^s]$ such that $i_2 \in [i_1 + 1, i_4]$; $j_2 \in [j_1 + 1, j_4]$; $i_3 \in [i_1, i_4 - 1]$; $j_3 \in [j_1, j_4 - 1]$; $z_{s, i_1, j_1} \ge z_{s, i_2, j_1}, z_{s, i_1, j_2}$; and $z_{s, i_4, j_4} \ge z_{s, i_3, j_4}, z_{s, i_4, j_3}$. Observe that the entries of $\textbf{Y}_{s - 1}$ decrease from top to bottom in any fixed column, decrease from left to right in any fixed row, and are less than $z_{s, i, j}$ for any $(i, j) \notin [2^{s - 1} + 1, 2^s] \times [2^{s - 1} + 1, 2^s]$. This implies that $(i_4, j_4) \notin [2^{s - 1} + 1, 2^s] \times [2^{s - 1} + 1, 2^s]$ and thus that either $i_4 \le 2^{s - 1}$ or $j_4 \le 2^{s - 1}$. We will only consider the case $i_4 \le 2^{s - 1}$ because the reasoning for the case $j_4 \le 2^{s - 1}$ is similar. It follows that $j_1 \le 2^{s - 1}$ or else $\textbf{Z}_{s - 1}$ would contain an acute angle matrix, contradicting the minimality of $s$; therefore, $(i_1, j_1), (i_2, j_1) \in [1, 2^{s - 1}] \times [1, 2^{s - 1}]$. However, the entries of any fixed column of $\textbf{X}_{s - 1}$ are increasing from top to bottom; this implies that $z_{s, i_1, j_1} < z_{s, i_2, j_1}$, which is a contradiction. Thus $\textbf{Z}_m$ has no acute angle submatrix for all positive integers $m$, so $\textbf{D}_m$ has the obtuse angle property. 

Next, suppose that there exists some positive integer $m$ such that $\textbf{D}_m$ does not satisfy the diagonal property; let $s \ge 2$ be the minimal such integer. Then there exist integers $i_1, j_1, i_2, j_2 \in [1, 2^s]$ with $i_1 < i_2$ and $j_1 < j_2$ such that $z_{s, i_1, j_1} + z_{s, i_2, j_2} \ge z_{s, i_2, j_1} + z_{s, i_1, j_2}$. First suppose that $j_2 \le 2^{s - 1}$ and $i_2 > 2^{s - 1}$ both hold. If we moreover had that $i_1 > 2^{s - 1}$, then the entries $z_{s, i_1, j_1}$, $z_{s, i_2, j_2}$, $z_{s, i_2, j_1}$, and $z_{s, i_1, j_2}$ would lie in the bottom-left $2^{s - 1} \times 2^{s - 1}$ corner of $\textbf{Z}_s$. Then $10^{-1000^s} \textbf{Z}_{s - 1}$ would not satisfy the diagonal property, which contradicts the minimality of $s$; hence $i_1 \le 2^{s - 1}$. 

Now, the difference between any two unequal entries of $\textbf{X}_{s - 1}$ has magnitude greater than the difference between any two entries of $10^{-1000^s} \textbf{Z}_{s - 1}$. Since the entries of $\textbf{X}_{s - 1}$ are increasing from left to right in any fixed row, this yields that $z_{s, i_1, j_2} - z_{s, i_1, j_1} = |z_{s, i_1, j_2} - z_{s, i_1, j_1}| > z_{s, i_2, j_2} - z_{s, i_2, j_1}$. This is a contradiction, which implies that either $j_2 > 2^{s - 1}$ or $i_2 \le 2^{s - 1}$. By similar reasoning, one may show that either $j_2 \le 2^{s - 1}$ or $i_2 > 2^{s - 1}$. It follows that $z_{s, i_2 j_2}$ is contained either in the top left $2^{s - 1} \times 2^{s - 1}$ corner of $\textbf{Z}_s$, which is a copy of $\textbf{X}_{s - 1}$, or in the bottom right $2^{s - 1} \times 2^{s - 1}$ corner of $\textbf{Z}_s$, which is a copy of $\textbf{Y}_{s - 1}$. By similar reasoning, one may deduce the same statement for $z_{s, i_1, j_1}$. 

Observe that any entry of $\textbf{Y}_{s - 1}$ is negative and has magnitude greater than $3$ times the magnitude of any entry of $\textbf{X}_{s - 1}$; furthermore, any entry of $\textbf{X}_{s - 1}$ has greater magnitude than any entry of $10^{-1000^s} \textbf{Z}_{s - 1}$. Hence if $z_{s, i_1, j_1}$ is in $\textbf{X}_{s - 1}$ and $z_{s, i_2, j_2}$ is in $\textbf{Y}_{s - 1}$, then $z_{s, i_1, j_1} + z_{s, i_2, j_2} < -2|z_{s, i_1, j_1}| < -|z_{s, i_2, j_1}| - |z_{s, i_1, j_2}| \le z_{s, i_2, j_1} + z_{s, i_1, j_2}$, which is a contradiction. Thus the entries $z_{s, i_1, j_1}$, $z_{s, i_2, j_2}$, $z_{s, i_2, j_1}$, and $z_{s, i_1, j_2}$ are either all in $\textbf{X}_{s - 1}$ or all in $\textbf{Y}_{s - 1}$. 

The inequality $i_1 j_1 + i_2 j_2 > i_1 j_2 + i_2 j_1$ implies that the latter case is impossible, so all four entries are in $\textbf{X}_{s - 1}$. If $i_1 + j_1 < 2^{s - 1} + 1$, then $z_{s, i_1, j_1}$ is negative and one may show that $|z_{s, i_1, j_1}| / 3 > |z_{s, i_2, j_2}|, |z_{s, i_1, j_2}|, |z_{s, i_2, j_1}|$. This implies that $z_{s, i_1, j_1} < -|z_{s, i_1, j_2}| - |z_{s, i_2, j_1}| - |z_{s, i_2, j_2}| \le z_{s, i_2, j_1} + z_{s, i_1, j_2} - z_{s, i_2, j_2}$, which is a contradiction. If $i_1 + j_1 = 2^{s - 1} + 1$, then $10^{1000^r} (z_{s, i_1, j_2} - z_{s, i_1, j_1}) \ge 1 > 10^{1000^r} (z_{s, i_2, j_2} - z_{s, i_2, j_1})$, which is a contradiction. Hence $i_1 + j_1 > 2^{s - 1} + 1$, so the inequality $(2^{s + j_1 - 1} - i_1)^2 + (2^{s + j_2 - 1} - i_2)^2 < (2^{s + j_1 - 1} - i_2)^2 + (2^{s + j_2 - 1} - i_1)^2$ implies that $z_{s, i_1, j_1} + z_{s, i_2, j_2} < z_{s, i_1, j_2} + z_{s, i_2, j_1}$. This is again a contradiction, which implies that $\textbf{D}_m$ has the diagonal property and is thus distance-like. \\

We conclude this article by asking whether there exists a distance-like matrix whose skeleton is forbidden. In view of Theorem 3, a negative answer implies the existence of a counterexample to the conjecture that $U_c (n) = \Theta (n)$. On the other hand, a positive answer may lead to a better understanding of unit distances between vertices of a convex polygon. 

\section{Acknowledgements}
The author heartily thanks J\'{a}nos Pach for his guidance; Joseph O'Rourke for his suggestions on publication; and Alok Aggarwal, Bal\'{a}zs Keszegh, Jacob Fox, Andrew Suk, and the referees for their valuable suggestions.

\end{document}